\documentclass[11pt,preprint]{aastex}
\usepackage{amsmath,amssymb,natbib,graphicx}

\newcommand{\icarus}{{\em Icarus}}

\newcommand{\g}{\mathrm{g}}

\newcommand{\cm}{\mathrm{cm}}

\newcommand{\au}{\mathrm{AU}}

\bibpunct{(}{)}{;}{a}{}{,}

\begin{document}
\title{Avoiding resonance capture in multi-planet extrasolar systems}

\author{Margaret~Pan\altaffilmark{1}, Hilke~E.~Schlichting\altaffilmark{1,2}}
\altaffiltext{1}{Massachusetts Institute of Technology, 77 Massachusetts Avenue, Cambridge, MA 02139}
\altaffiltext{2}{University of California, Los Angeles, 595 Charles E. Young Drive East, Los Angeles, CA 90095}

\begin{abstract}
  A commonly noted feature of the population of multi-planet
  extrasolar systems is the rarity of planet pairs in low-order
  mean-motion resonances.  We revisit the physics of resonance capture
  via convergent disk-driven migration. We point out that for planet
  spacings typical of stable configurations for {\em Kepler} systems,
  the planets can routinely maintain a small but nonzero eccentricity
  due to gravitational perturbations from their neighbors. Together
  with the upper limit on the migration rate needed for capture, the
  finite eccentricity can make resonance capture difficult or
  impossible in Sun-like systems for planets smaller than
  $\sim$Neptune-sized. This mass limit on efficient capture is broadly
  consistent with observed exoplanet pairs that have mass
  determinations: of pairs with the heavier planet exterior to the
  lighter planet --- which would have been undergoing convergent
  migration in their disks --- those in or nearly in resonance are
  much more likely to have total mass greater than two Neptune masses
  than to have smaller masses. The agreement suggests that the
  observed paucity of resonant pairs around sun-like stars may simply
  arise from a small resonance capture probability for lower-mass
  planets. Planet pairs that thereby avoid resonance capture are much less
  likely to collide in an eventual close approach than to simply
  migrate past one another to become a divergently migrating pair with
  the lighter planet exterior. For systems around M stars we expect
  resonant pairs to be much more common, since there the minimum mass
  threshhold for efficient capture is about an Earth mass.
  \end{abstract}

\section{Introduction}

In the last decade several hundred multiple-planet systems, together
containing over 1300 planets, have been discovered, mostly through
transit photometry surveys such as the {\em Kepler} mission. These
systems comprise a statistically interesting sample of exoplanetary
orbital configurations that can constrain the roles of different
planet-planet interactions during system formation. 

\begin{figure}
\begin{center}
  \includegraphics[scale=.7]{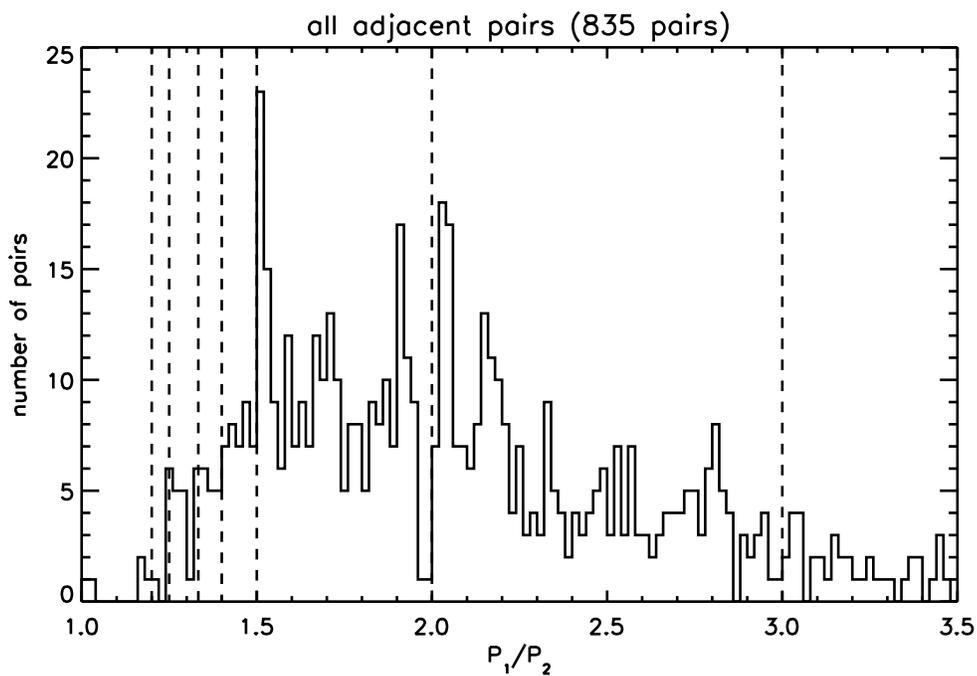}
  \caption{Period ratio histogram of all pairs of adjacent planets in
    the NASA Exoplanet Archive as of 2016 May 25. While peaks in the
    histogram occur just outside several mean motion resonances
    (dashed lines), most pairs seem to be completely unassociated with
    such resonances. \label{fig:ratiosall}}
\end{center}
\end{figure}

Of particular interest here is the occurrence of (near-)mean-motion
resonances among planet pairs. Studies of the {\em Kepler} planet
population indicate that these bodies are typically between Earth- and
Neptune-sized and that typically $\sim$1\% of the planet mass lies in
a hydrogen/helium envelope \citep[see, for example,][and references
  therein]{wolfgang15}. The presence of these light gases strongly
suggests that these planets formed while their protoplanetary disks
were still present. In the standard example of a minimum mass solar
nebula (MMSN)-like disk, we expect embedded planets to migrate inwards
towards the star at speeds proportional to their masses. For Earth- to
Jupiter-mass planets with very low eccentricities, pairs of planets
undergoing convergent migration have an order unity chance of becoming
locked in first-order resonances when their semimajor axis ratio
reaches the relevant value. Nonetheless, a glance at the period ratios of
adjacent planet pairs (pairs not known to have additional planets
separating them) indicates that at most a few percent are currently
likely to be in resonance \citep[see Figure~\ref{fig:ratiosall}
  and][]{fabrycky14,steffen15}. Pairs with period ratios just larger
than several first-order mean motion resonance values do appear to
occur more frequently than a smooth distribution would predict. This
suggests that the nearby resonances may have significantly affected
those pairs' dynamics, and several groups have offered explanations
for the observed offsets \citep[see, for example,][and references
  therein]{lithwick12,batygin13,baruteau13,petrovich13,delisle14,chatterjee15}.

However, even these near-resonant pairs are only $\sim$15\% of the
total (see Figure~\ref{fig:ratiosall}), not enough to be consistent
with the large nominal capture probability.  To explain their rarity,
several groups \citep[see, for example,][]{adams08,rein09} have
proposed that turbulence may excite resonant planets' random motions
enough to disrupt the resonance.  \citet{goldreich14} and
\citet{delisle14} proposed that pairs initially trapped in resonance
might later escape due to overstable librations coupled with
eccentricity damping. After escape, the planets would migrate away
from resonance. On the other hand, \citet{batygin15} proposed that the
embedded planets' eccentricities might be too large for efficient
resonant capture to occur at all despite convergent migration. He
finds $e\gtrsim 10^{-2}$ as the eccentricity criterion for efficient
capture and notes that it is comparable to the typical Kepler planet
eccentricity \citep{wu13,hadden14}. However, he does not address these
finite eccentricities' origin, or whether they occur during or after
the time when the disk is present and migration is expected.

Here we discuss an explanation for why $e\gtrsim 10^{-2}$ may
typically occur for many planets during disk migration: gravitational
interactions between planets as they move through the disk. For
smaller planets separated by several Hill spheres, close encounters
between neighboring planets can excite eccentricities faster than
planet-disk interactions can damp them. Combined with the requirement
for efficient resonance capture that disk migration be slower than the
resonant libration period, these finite eccentricities significantly
limit the range of planet masses that can routinely capture into
resonance. In \S\ref{sec:setup} we define system parameters; in
\S\ref{sec:timescales} we discuss timescales for migration and close
encounters; in \S\ref{sec:eccentricity} we discuss the typical planet
eccentricity; in \S\ref{sec:capture} we discuss the implications for
resonant capture in the context of the known multiplanet systems; and
in \S\ref{sec:summary} we summarize our findings.

\section{Planetary system parameters}\label{sec:setup}

We assume a young solar system centered on a star of mass $M_*$,
radius $R_*$, and bulk density $\rho_*$. The system contains a
passively heated mostly gaseous circumstellar disk similar to the
minimum mass solar nebula (MMSN). The disk has mass surface density
$\sigma_g(a)\propto a^{-3/2}$ (where $a$ is the semimajor axis); scale
height $h_g\propto a^{5/4}$; and volume density
$\rho_g\sim\sigma_g/h_g$. Embedded in the disk are young planets near
the isolation mass \citep{lissauer87}. Their orbital spacings are typically of order a
few times their Hill spheres, $R_H\simeq \mu^{1/3}a$, where $\mu$ is
the planet/star mass ratio, $\mu M_*$ is the planet mass, and $a$ is
the planet semimajor axis.  In discussions of rates of collision and
resonant capture we focus on two planets of masses $\mu M_*$ and
$f^3\mu M_*$ where $f$ is a numeric constant; bulk density $\rho$; and
sizes $R$ and $fR$. The planets orbit at semimajor axes $a_1$, $a_2$
and with eccentricities $e_1$, $e_2$. We assume the planets are too
small to open gaps in the disk. We also assume $\rho_*\simeq\rho$; for
a late main-sequence star and typical planets, both are of order unity
($\g\,\cm^{-3}$).

For convenience, we define $\alpha = R_*/a$, so that the Hill radius
becomes $R_H = \alpha^{-1} R$ and Hill velocity is
$v_H=\alpha^{1/2}v_\mathrm{esc}\simeq\alpha^{1/2}R\sqrt{G\rho}$.  We
also use the usual orbital angular frequency
$\Omega\simeq\sqrt{G\rho}\alpha^{3/2}$, orbital velocity
$v_K\simeq\Omega a$, and gas sound speed $c_g\simeq h_g\Omega$.

If migration is integral to planetary system architectures, we would
expect planets currently at or inside $a=1$~AU to have formed and
interacted further out.  For the sake of representative numerical
estimates, unless otherwise noted, we assume a MMSN disk around a
solar type star at a semimajor axis 10 AU. These parameters imply
$\alpha\sim 0.0005$, $\sigma_g\sim 55\,\g\,\cm^{-2}$, $h_g/a\sim
0.18$, which we will refer to as ``standard conditions''. They
give an isolation planet mass of $\sim$$4\times 10^{27}$~g \citep[see, for example,][]{schlichting14}.

\section{Encounter timescales\label{sec:timescales}}

\subsection{Migration}\label{sec:migration}

Convergent disk-mediated migration is a natural mechanism for inducing
planet pairs' resonance crossings and possible capture. We briefly review
migration caused by gas drag and by gravitational torques exerted by
the disk on the planet.

Due to pressure support, the gas orbits slower than the local Keplerian velocity by a fraction
\begin{equation}
  \frac{\delta v}{v_K}\sim \left(\frac{c_g}{v_K}\right)^2 \;\;\;.
  \label{eq:deltav}
\end{equation}
As a result planets moving through the gas feel a drag force and drift towards the star at a rate
\begin{equation}
  \frac{1}{a}\frac{da}{dt} \sim \frac{\sigma_g}{\rho R}\frac{h_g^3}{a^3}\Omega
  \sim \frac{\mu_d}{\mu^{1/3}}\alpha^2\frac{h_g^3}{a^3}\Omega \;\;\;,
  \label{eq:stokestime}
\end{equation}
assuming they are in the Stokes regime. Here, $\mu_d\sim\sigma_g
a^2/M_*$ is the ratio of the stellar mass and the local disk mass, or
the disk mass within a semimajor axis range of about a factor of two
about $a$; it does not represent the total disk mass. At the same
time, gravitational interactions between a planet and nearby disk gas
lead to torque exchange. In a circular disk where the density
decreases sufficiently slowly outwards, and where the viscosity
circularizes gas molecules' trajectories within one synodic period,
the asymmetry between interactions with interior and exterior gas is
of order $h_g/a$, giving inward Type I migration rate
\begin{equation}
  \frac{1}{a}\frac{da}{dt}\sim \frac{\mu^2}{(h_g/a)^5}\cdot \frac{\sigma_g\,h_g/a}{\rho R^3}\cdot \frac{h_g}{a} \cdot \frac{h_g}{a}\Omega
  \sim \mu\mu_d\left(\frac{a}{h_g}\right)^2\Omega
  \;\;\;.
  \label{eqn:type1time}
\end{equation}

\subsection{Encounter timescales}\label{sec:enctime}

Two planets may capture into resonance only while their semimajor axis
ratio differs from exact resonance by less than the (fractional)
resonance width $\sim$$\mu^{2/3}$ \citep{murray99}. We take an
encounter to last while the semimajor axes of the planets in question
are within this range. If the planets are migrating due to disk
interactions as described above, the time to cross this distance is
\begin{equation}
  \mu^{2/3}\left(\frac{1}{a}\frac{da}{dt}\right)^{-1}
  \sim \left\{
  \begin{aligned}
    &\frac{\mu}{\mu_d}\frac{a^3}{h_g^3}\alpha^{-2}\frac{|1-f^{-1}|^{-1}}{\Omega} \propto a^{9/4}\mu^1 \qquad &\mathrm{Stokes}\\
    &\frac{1}{\mu_d\mu^{1/3}}\left(\frac{h_g}{a}\right)^2\frac{|1-f^3|^{-1}}{\Omega} \propto a^{3/2}\mu^{-1/3} &\mathrm{Type\;I}
  \end{aligned}\right.
  \label{eq:crosstime}
\end{equation}
where $\mu$ is the larger of $\mu_1$, $\mu_2$.  Note the Stokes drag
timescale increases while the Type I timescale decreases with
increasing $\mu$. The two inward migration rates match at
\begin{equation}
\mu\sim \alpha^{3/2}\left(\frac{h_g}{a}\right)^{15/4} \propto a^{-9/16}
\;\;\;,
\label{eq:slowestr}
\end{equation}
which corresponds to an upper bound on the encounter time of
\begin{equation}
\sim \frac{1}{\mu_d}\alpha^{-1/2}\left(\frac{h_g}{a}\right)^{3/4}\frac{1}{\Omega} \propto a^{27/16}\;\;\;.
\end{equation}
For a MMSN around a solar type star at 10~AU, the slowest encounter
takes $\sim$$10^5$~yrs and occurs for $(\mu_1, \mu_2)\sim \mu\sim
10^{-8}$, or $R\sim 10^3$~km.

\section{Typical eccentricities\label{sec:eccentricity}}

Perhaps the physically simplest way to excite planets' eccentricities
is gravitational perturbations from neighboring bodies. To find the
typical eccentricity thus induced, we first calculate the
gravitational stirring rate. We take the typical spacing between
planets to be $CR_H$ where $C\simeq 10$, following the multiplanet
stability limit simulations of \citet{pu15}. Since planet formation
theory also predicts spacings of $\sim$several $R_H$ during oligarchy
\citep[see, for example,][and references therein]{goldreich04}, we
believe this is a reasonable estimate for planet separations. A close
approach within $R_H$ typically gives the planets an additional random
velocity of $v_H$, or eccentricity $v_H/(\Omega a)\sim
\mu^{1/3}$. Then bodies of mass $\mu$ spaced $CR_H$ apart typically
impart eccentricities $\sim \mu^{1/3}C^{-2}$ to one another during a
single conjunction\footnote{This approximation is reasonable as long
  as $CR_H$ is significantly less than unity. E.g. with $C=10$, this
  holds for planets significantly smaller than a Jupiter mass. For
  larger separations, it is a slight underestimate: at the separation
  of bodies in the 2:1 resonance, the inverse square scaling with
  distance gives an interaction too small by a factor of 1.4.}.

This eccentricity is damped by disk interactions at a rate
\begin{equation}
\frac{1}{e}\frac{de}{dt}\sim \mu\mu_d\left(\frac{a}{h_g}\right)^4\Omega \propto a^{-2}\mu^1
\end{equation}
per \citet{goldreich14}. This is slower than the synodic
frequency $\Omega CR_H/a$ as long as
\begin{equation}
\mu\lesssim C^{3/2}\mu_d^{-3/2}\left(\frac{h_g}{a}\right)^6 \;\;\;,
\end{equation}
which is easily satisfied for all the cases we consider here. This
implies that the eccentricity gained at one conjunction will not be
damped before the next conjunction. Since the planets are not in
resonance, the longitudes of successive conjunctions are uncorrelated,
and the planets' eccentricities increase per a random walk with steps
of size $\sim \mu^{1/3}C^{-2}$ taken once every synodic time until an
eccentricity damping time has elapsed. This yields
\begin{equation}
  \text{typical eccentricity}\sim C^{-3/2}\mu_d^{-1/2}\left(\frac{h_g}{a}\right)^2\;\;\;.
  \label{eqn:etyp}
\end{equation}
Note that the typical eccentricity is independent of $\mu$.

Finally, even if the typical separation is $CR_H$, planets migrating
at different rates will occasionally pass one another, leading to a
close approach within $R_H$. As we discuss in \S\ref{sec:location},
direct collisions are unlikely during these close approaches. However,
if the larger eccentricities attained at these $\lesssim R_H$
approaches persist at larger separations, they may dominate the
typical excitation from stirring expressed in
Equation~\ref{eqn:etyp}. We therefore check whether eccentricities
excited during these very close approaches can damp quickly as the
neighboring planets differentially migrate away from each other. To do
this we compare the eccentricity damping timescale to the time
required to migrate through $R_H$:
\begin{equation}
  \frac{\left.\dfrac{1}{e}\dfrac{de}{dt}\right|_\mathrm{damp}}{\dfrac{1}{\mu^{1/3}a}\dfrac{da}{dt}}\sim\left\{
  \begin{aligned}
    &\mu^{5/3}\left(\frac{a}{h_g}\right)^7\alpha^{-2}\;&\text{Stokes}\\
    &\mu^{1/3}\left(\frac{a}{h_g}\right)^2\;&\text{Type I}
\end{aligned}\right.\;\;\;.
\end{equation}
Setting the above to unity and solving for $\mu$, we find
\begin{equation}
  \mu \gtrsim\left\{
  \begin{aligned}
    &\left(\frac{h_g}{a}\right)^6 \propto a^{3/2}\;&\text{Stokes}\\
    &\left(\frac{h_g}{a}\right)^{21/5}\alpha^{6/5}\propto a^{-3/20}\;&\text{Type I}
  \end{aligned}\right.
\end{equation}
is required for damping to occur faster than migration through $R_H$.
For our standard disk conditions, eccentricity damping occurs faster
than migration through $R_H$ only for planets larger than $\mu\sim
9\times 10^{-7}$. For smaller planets down to the Type I-Stokes
boundary, eccentricities of order $e_\mathrm{crit}$ persist as the
planets migrate past nearest-neighbor separation $R_H$. Thus
for planets larger than $\mu\sim (h_g/a)^{21/5}\alpha^{6/5}$,
eccentricities are excited to $\gtrsim$$e_\mathrm{crit}$ while their
nearest neighbor planet orbit lies within $R_H$; once they migrate
away, their eccentricities settle to a new equilibrium set by more
distant close approaches at $\sim CR_H$. Equation~\ref{eqn:etyp}
accurately represents these planets' typical eccentricities. For
smaller planets, eccentricities grow to $\gtrsim$$e_\mathrm{crit}$
while the nearest neighbor planet orbit lies within $R_H$ and do not
damp quickly afterwards. For much of the time, these planets'
eccentricities may be larger than that given by
Equation~\ref{eqn:etyp}.

\section{Resonance capture}\label{sec:capture}

\subsection{Chance of capture per encounter}\label{sec:rescapt}

Two convergently migrating planets may eventually reach orbital
separations consistent with a $p:q$ mean-motion resonance, that is, a
semi-major axis ratio closer to the exact resonance ratio
$(p/q)^{2/3}$ than the fractional resonance width $\sim$$\mu^{2/3}$.
To guarantee resonant capture during the planets' pass through this
range, the time to cross the resonance must be longer than the
critical libration timescale $\sim$$\mu^{-2/3}\Omega^{-1}$
\citep{goldreich14}. This requires
\begin{equation}
  \mu \gtrsim \left\{
  \begin{aligned}
    \alpha^{6/5}\mu_d^{3/5}\left(\frac{h_g}{a}\right)^{9/5} &\propto a^{-9/20}\;& \text{Stokes}\\
    \mu_d^3\left(\frac{a}{h_g}\right)^6 &\propto a^{-1} &\text{Type I}
  \end{aligned}\right.\;\;\;.
  \label{eqn:rescapt}
\end{equation}

However, this argument only makes sense if, despite conjunctions with
nearby planets before an encounter, the eccentricities of the planets
undergoing the encounter remain low enough to make capture likely. The
eccentricity threshhold for efficient resonance capture is
$e_\mathrm{crit}\sim \mu^{1/3}$ \citep{goldreich14}, so planets whose
typical eccentricity remains below $e_\mathrm{crit}$ should capture
efficiently. Using Equation~\ref{eqn:etyp}, these planets must have
\begin{equation}
  \mu\gtrsim C^{-9/2}\mu_d^{-3/2}\left(\frac{h_g}{a}\right)^6\propto a^{3/4} \;\;\;.
  \label{eqn:eccmass}
\end{equation}
We would expect smaller planets to have typical eccentricity
larger than $\mu^{2/3}$. At those higher eccentricities, capture is
possible but less likely: the eccentricities of planets librating in
resonance may oscillate through larger values, but in order for
capture to occur at high eccentricities, the phases of the planets as
they reach the resonance must match the phases where high
eccentricities are attained. Using the Hamiltonian as in
\citet{murray99}, we calculate the capture probability to decrease as
$[\text{eccentricity}]^{-1.43}$.

\subsection{Resonant planet pairs}\label{sec:respairs}

\subsubsection{Sun-like stars}

\begin{figure}
  \begin{center}
    \includegraphics[scale=.45]{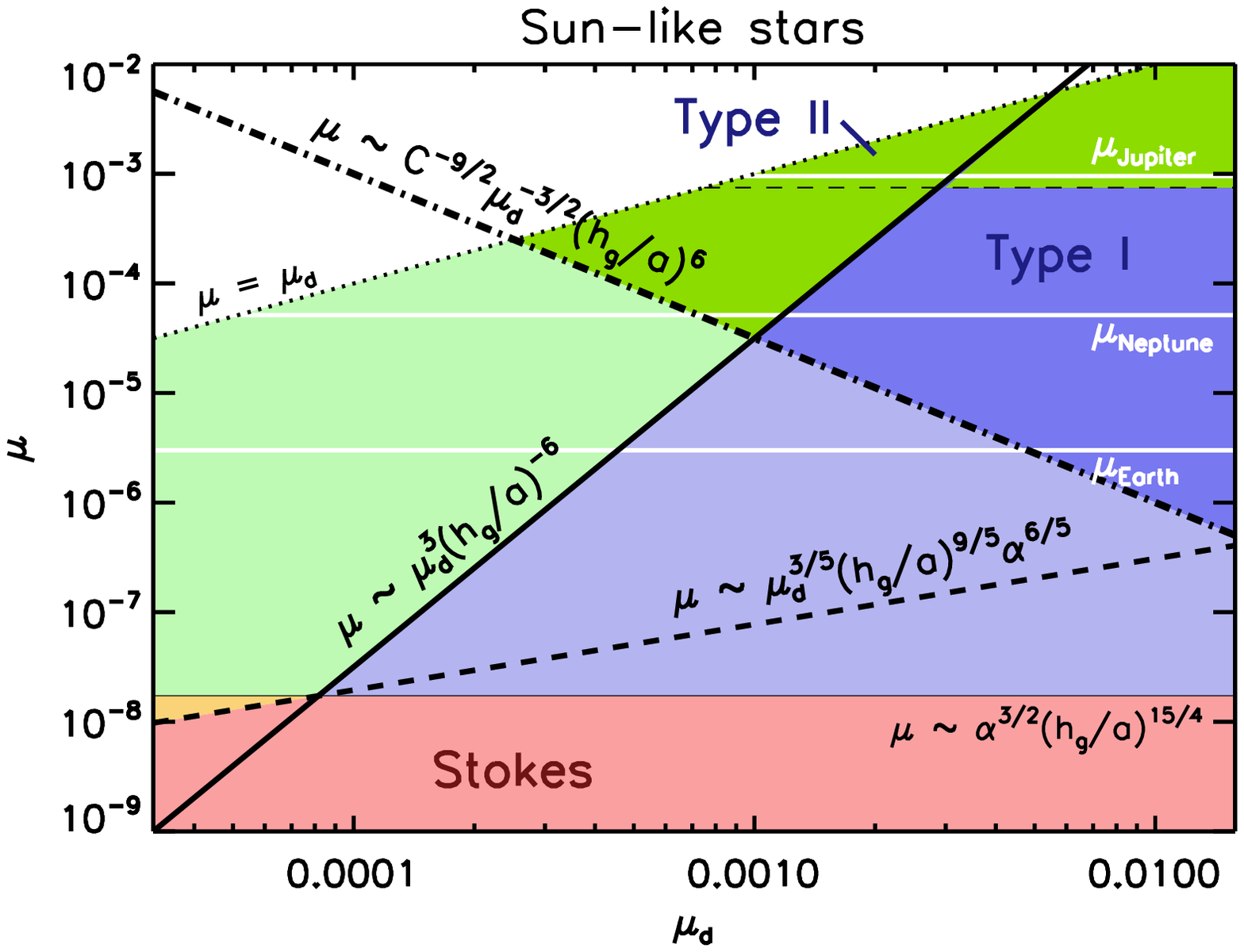}
\includegraphics[scale=.45]{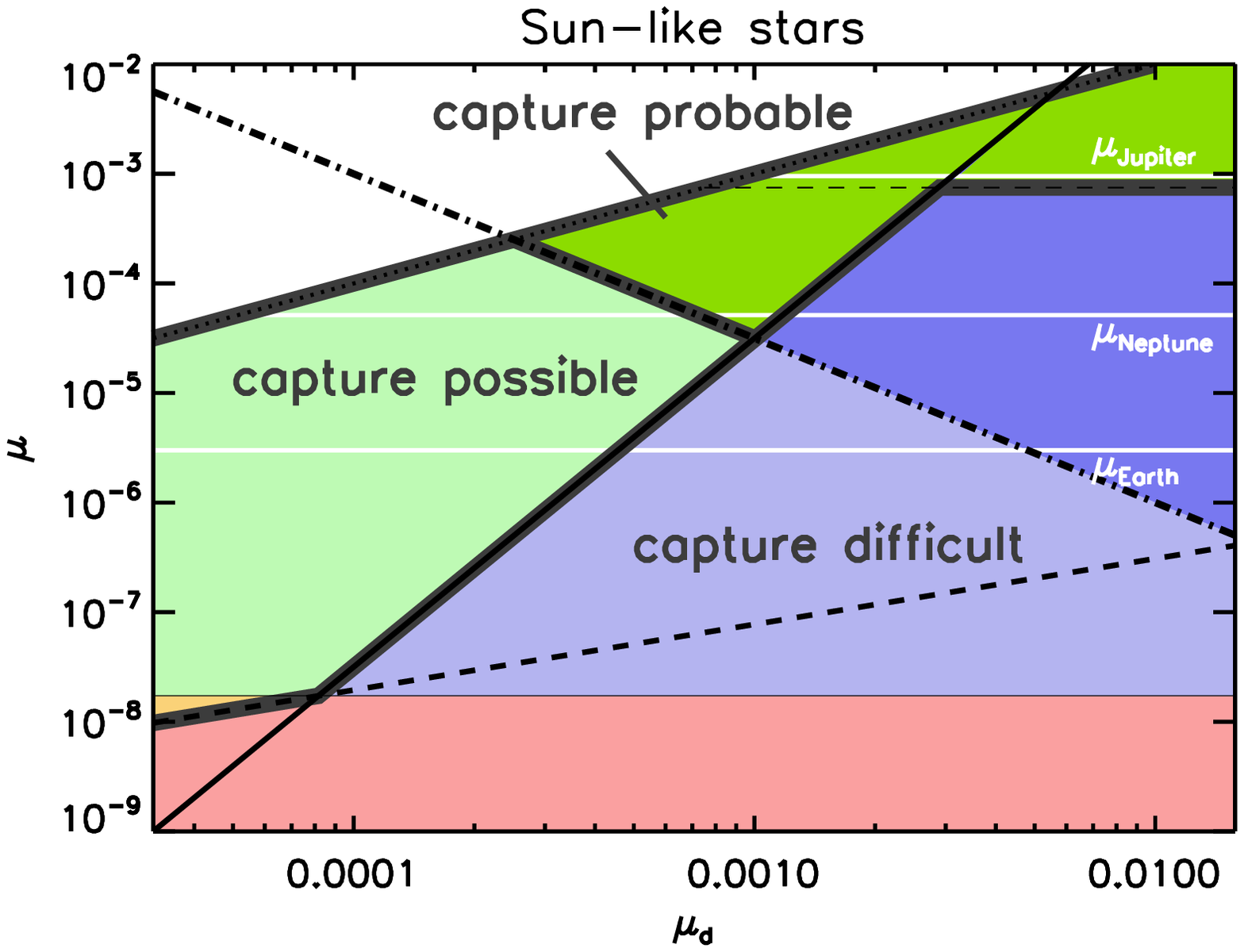}
    \includegraphics[scale=.45]{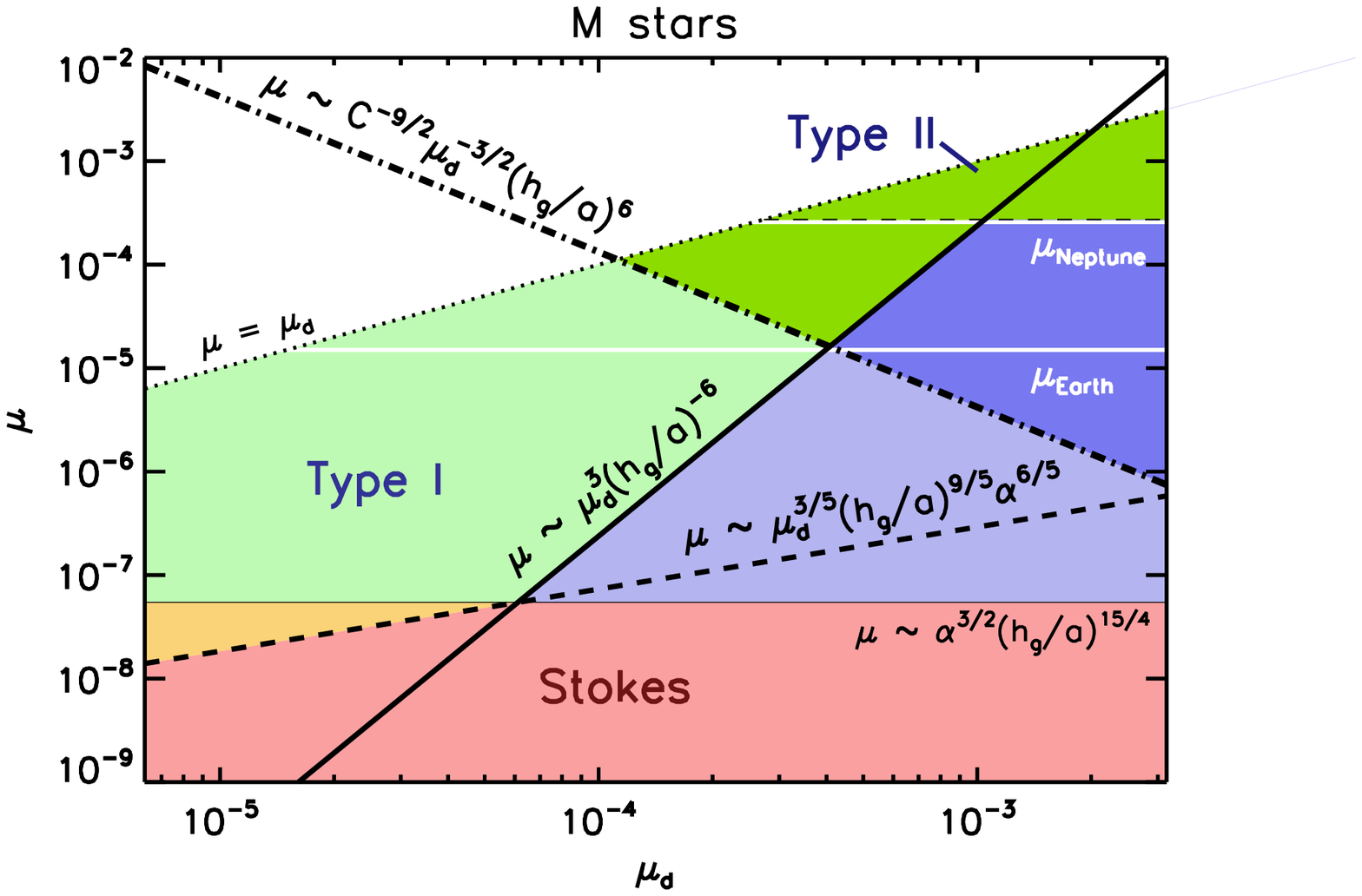}
    \includegraphics[scale=.45]{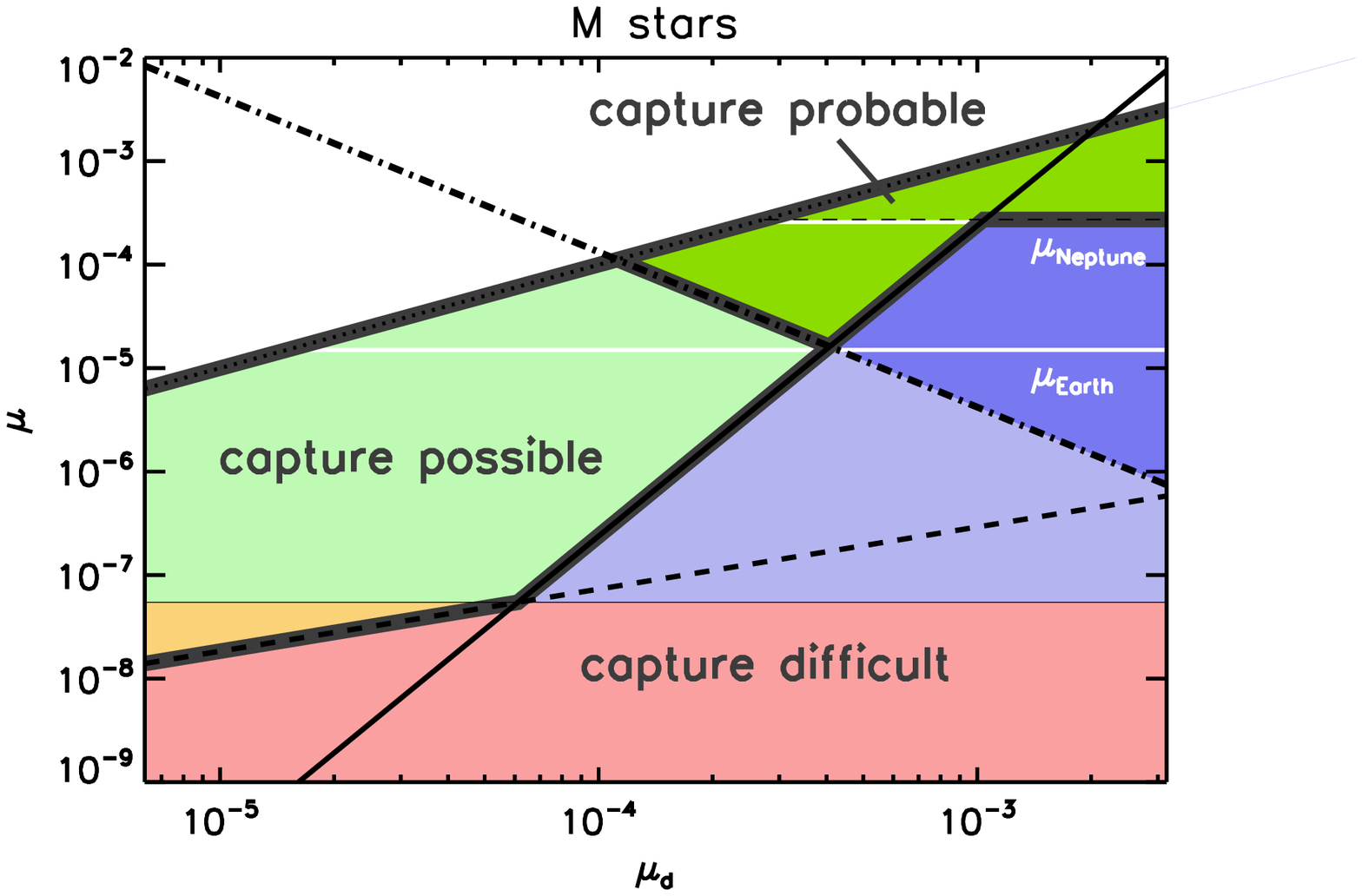}
    \caption{Summary plots showing areas of likely resonant capture
      (bright green), possible resonant capture (light green, orange),
      and unlikely resonant capture (blue, red) as a function of
      planet mass ratio $\mu$ and disk mass ratio $\mu_d$. Left and
      right hand columns show the same results, but plots in the
      right column emphasize regions corresponding to different
      resonant encounter outcomes. Top panel assumes a star of mass
      $M_\odot$ and radius $R_\odot$, semimajor axis 10~AU, and disk
      scale height $h_g/a=0.1(a/1\,\au)^{1/4}$; bottom panel assumes a
      star of mass $M_\odot/4$ and radius $R_\odot/2$, semimajor axis
      1~AU, and disk scale height $h_g/a=0.1(a/0.4\,\au)^{1/4}$. In
      red/orange regions, migration by Stokes drag dominates; in
      blue/green regions, Type I migration is faster. The most massive
      planets (above the thin dashed line; numbers from Figure~6 of
      \citet{rafikov02} assuming disk $\alpha$-parameter
      $\sim$$10^{-5}$) open gaps in their disks, slowing their
      migration dramatically. In regions of unlikely resonant capture,
      the time for migration across a resonance width is faster than
      the libration period. In regions of likely capture, migration is
      sufficiently slow and typical eccentricities are sufficiently
      small for efficient capture. In regions of possible capture,
      migration is sufficiently slow but eccentricities are larger
      than the threshhold value $\sim$$\mu^{1/3}$; here the
      probability of capture decreases approximately as
      $[\text{eccentricity}]^{-1.43}$. Since it is unrealistic for the
      planets to be more massive than their disk at this stage, all
      regions are cut off at $\mu=\mu_d$.  For sun-like systems, only
      planets more massive than about a Neptune mass are likely to be
      captured. For M dwarf systems, planets larger than about half an
      Earth mass are likely to be captured. In both cases a
      sufficiently massive disk is required; for sun-like systems the
      lower limit for efficient capture is close to the MMSN, and for
      M dwarf systems it is close to a $0.01M_*$ disk covering the
      range 0.01 to 5 AU with $\sigma_g\propto
      a^{-3/2}$.\label{fig:rates}}
  \end{center}
\end{figure}

For solar system-like conditions, that is, $\mu_d\sim 10^{-3}$ at
10~AU, the numerical estimates in Sections~\ref{sec:eccentricity} and
\ref{sec:rescapt} apply: planets with $\mu\gtrsim 3\times 10^{-5}$
migrate slowly enough and maintain sufficiently small eccentricities
to capture into resonance during a close encounter. For disk masses
smaller than this, planet-planet scattering increases the
eccentricities of all planets above $e_\mathrm{crit}$ before resonance
entry, greatly decreasing the chance of capture. Occasional close
approaches within $R_H$ do not strongly affect the typical
eccentricity for planets larger than $\mu\sim8\times 10^{-8}$. For
larger disk masses, the smallest planet that migrates slowly enough to
be captured increases as $\mu_d^3$ until that planet becomes large
enough to open a gap in the disk. Once this occurs, the planet becomes
locked to the disk and its migration rate slows dramatically. The
planet mass where this occurs depends on the disk viscosity; as a
representative value we take $\mu\sim 0.2(h_g/a)^3$ using Figure~6 of
\citet{rafikov02}. The above results are summarized in the top panels of
Figure~\ref{fig:rates}.

\begin{figure}
\begin{center}
\includegraphics[scale=.7]{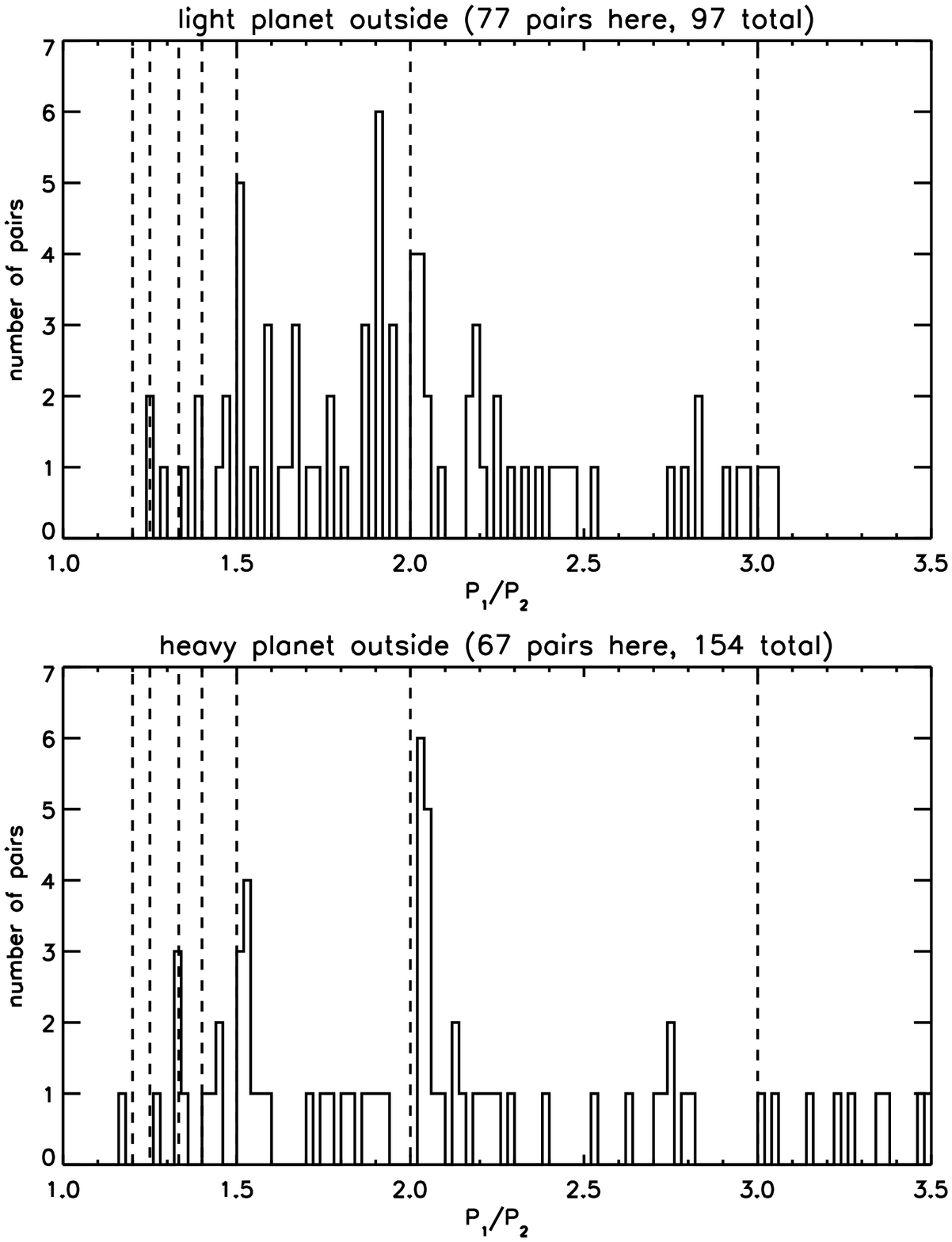}
\caption{Histograms of orbit period ratios for pairs of adjacent
  Kepler planets (pairs not known to have additional planets between
  them) in multi-planet systems with measured masses as of 2016 May
  25. Mean-motion resonance locations are indicated by dashed
  lines. When the outer planet is the more massive of the pair (bottom
  panel), the peaks near resonances are much more pronounced than when
  the outer planet is the less massive (top panel). Since convergent
  migration in a MMSN-type disk occurs only when the more massive
  planet is exterior, this supports the idea that migration indeed
  pushes planet pairs into resonance. \label{fig:ratios} }
\end{center}
\end{figure}

\begin{figure}
\begin{center}
  \includegraphics[scale=.7]{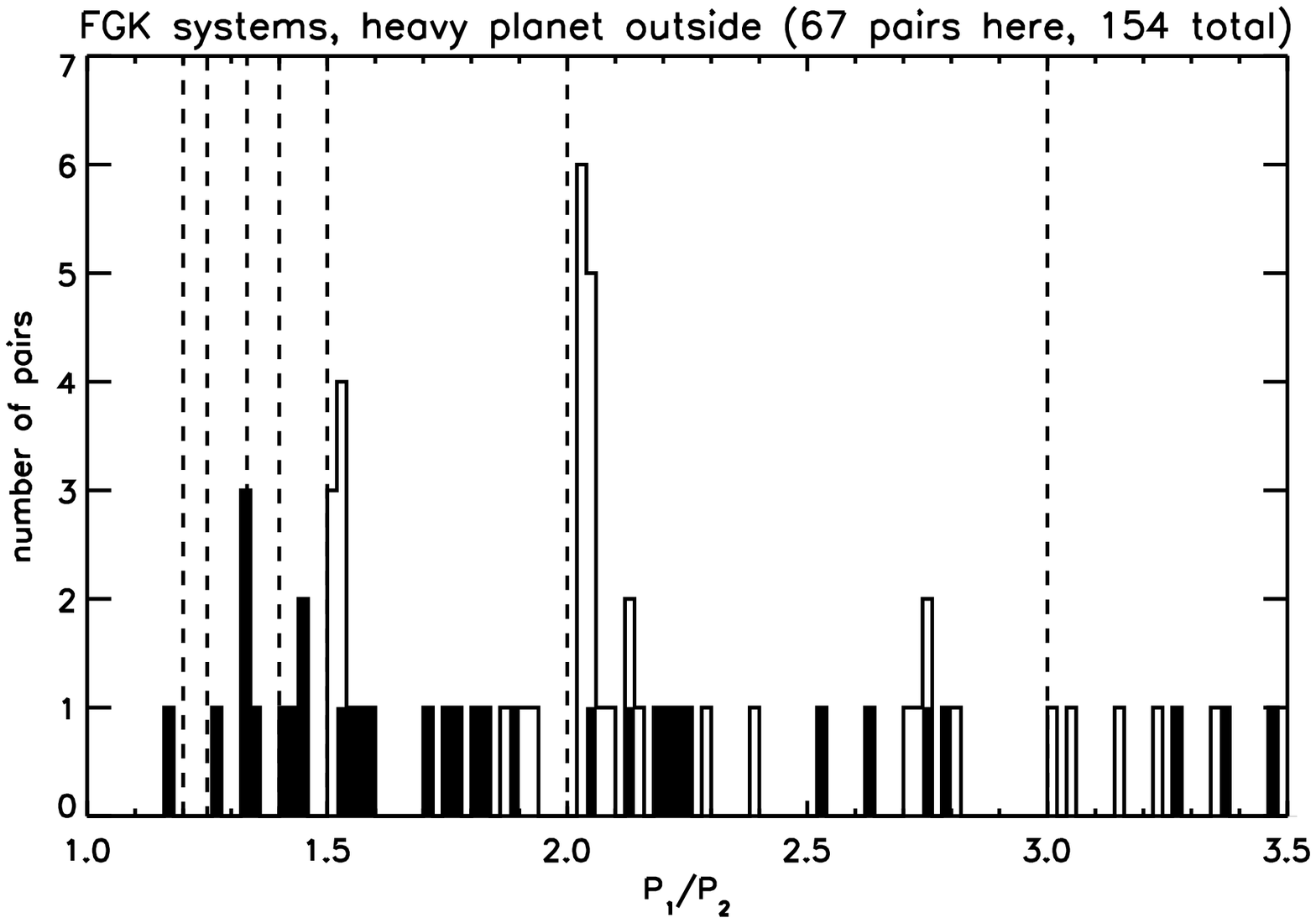}
  \caption{Similar to the bottom panel of Figure~\ref{fig:ratios}
    except that while the unshaded histogram includes all such pairs,
    the shaded histogram includes all pairs with combined mass less
    than $0.1M_\mathrm{Jupiter}\simeq 2M_\mathrm{Neptune}$. That the
    vast majority of near-resonant pairs occur among systems with
    combined planet mass $>2M_\mathrm{Neptune}$ supports our
    estimates in Section~\ref{sec:respairs}. \label{fig:ratiosb}}
\end{center}
\end{figure}

The Kepler planet sample is in broad agreement with this estimate
(Figures~\ref{fig:ratios} and \ref{fig:ratiosb}): among pairs of
adjacent planets with the more massive planet outside --- a
configuration consistent with convergent migration in a MMSN-type disk
--- near-resonant pairs are much more likely to have combined mass
$>$2~Neptune masses than to have lower masses. The 2:1, 3:2, and 4:3
resonances are associated with 18 higher mass pairs of combined mass
$>2M_\mathrm{Neptune}$ vs. 6 lower mass pairs of combined mass
$<2M_\mathrm{Neptune}$. Of those 6, one includes a planet of mass
$1.2M_\mathrm{Neptune}$ and one orbits an M dwarf. At the same time,
the period distribution shapes in the top (lighter planet exterior)
and bottom (heavier planet exterior) panels of Figure~\ref{fig:ratios}
differ qualitatively: the near-resonance peaks appear much more
pronounced in the bottom than in the top panel. This supports the idea
that convergent migration leading to resonance capture does occur
among exoplanets.

\subsubsection{M stars}

Because $\sim$Earth-sized planets in the habitable zones of M stars
produce deeper transits more frequently than their analogues orbiting
solar-type stars, interest in M star planetary systems continues to
grow. An early- to mid-M main-sequence star might have $M_*=M_\odot/5$,
$R=R_\odot/2$, planets near $a=0.1$~AU, and a disk mass of $0.01M_*$
spread between 0.01~AU and 5~AU. As with the sun-like systems, we
assume the planets formed and interacted further out, so we consider
the dynamics at 1~AU. With $\sigma_g\propto a^{-3/2}$, we have
$\alpha\sim 0.0025$, $\sigma_g\sim 3.3\times 10^2$, and $h_g/a\sim
0.13$. Results for these system parameters are summarized in the
bottom panels of Figure~\ref{fig:rates}: we expect M dwarf planets in
the Earth- to super-Earth size range and larger to be able to capture
efficiently into resonance. Although the number of M dwarf systems
with measured planet masses is currently too small to provide
meaningful statistical contrast with the sample of FGK systems, we
would expect future surveys such as the {\em TESS} mission to show
similar period ratio distributions roughly independent of planet
mass. Nonetheless, systems such as TRAPPIST-1, with seven planets all
below 1.5$M_\mathrm{Earth}$ in resonances \citep{gillon17}, and
Kepler-32, with three out of five planets near mean-motion resonances
and smaller than 0.6~Neptune radii \citep{fabrycky12}, appear broadly
consistent with our prediction.

\subsubsection{Location of resonant encounters and close approaches\label{sec:location}}

In the above discussions we assumed the resonant interactions occurred
at significantly larger semimajor axes than those common in known
multiplanet systems. Taking the resonant encounters to have occurred
at the planets' current locations instead would require migration to
have stopped immediately post-encounter, and we have no reason to
expect such fine-tuned agreement between the disk lifetime
and the time of the encounter. However, from Equations~\ref{eqn:rescapt} and \ref{eqn:eccmass}, the minimum planet mass that efficiently captures
into resonance\footnote{As is true in the sun-like and M star systems
  discussed here, we assume planets subject to efficient resonance
  capture are in the Type I regime.},
\begin{equation}
  \mu_\mathrm{min}\sim C^{-3}\left(\frac{h_g}{a}\right)^2\propto a^{1/2} \;\;\;,
\label{eqn:mumin}
\end{equation}
decreases with decreasing semimajor axis. For example, taking our
usual MMSN-like system parameters at 1~AU with $C=10$ gives
$\mu_\mathrm{min}\sim 10^{-5}$, equivalent to
$\sim$$3M_\mathrm{Earth}$, rather than the
$\mu_\mathrm{min}\sim3\times 10^{-5}$ we find at 10~AU. The reasonable
agreement between our numbers (Figure~\ref{fig:rates}) and observed
systems (Figure~\ref{fig:ratiosb}) suggests that most resonant
encounters occurred in disks with $\mu_d$ and $h_g/a$ similar to those
of our standard conditions. In a MMSN-like system, this implies
semimajor axis values beyond the ice line. Planet pairs that failed to
capture into resonance via convergent migration at those larger
semimajor axes would eventually have a close approach, stir one
another, and continue inwards, most likely migrating
divergently. Planet-planet collisions during these close passages are
unlikely: the chance of collision is roughly the collision probability
per conjunction times the number of conjunctions that occur as the
planets migrate through their mutual Hill sphere,
\begin{equation}
  \alpha \cdot \frac{\mu^{2/3}\mu_d\dfrac{a^2}{h_g^2}\Omega}{\mu^{1/3}\Omega}
  \sim \mu^{1/3}\mu_d\left(\frac{a}{h_g}\right)^2\alpha
  \propto \mu^{1/3}a^{-1} \;\;\;,
\end{equation}
where we applied Equation~\ref{eqn:type1time}.  Here we assumed that
the planets undergo Type I migration and that their random velocities
are $\sim$$v_H$, since it takes just a few conjunctions within $R_H$
for them to stir each other's velocities to this level
\citep{goldreich04}. With our standard disk parameters and planets of
$\sim$5~Earth masses, the chance of collision is $\sim$$10^{-7}$. This
suggests that many pairs that today have the lighter planet exterior
could have originally been convergently migrating with the heavier
planet exterior.

Finally, when $([(p+1)/p]^{2/3}-1)\mu^{-1/3}\lesssim 10$, the spacing
between two planets at the $p+1:p$ resonance is such that
$C<10$. Since resonance capture would for these larger planets require
a nearest neighbor closer than $C=10$, we would expect them to acquire
eccentricities larger than that given by Equation~\ref{eqn:etyp} with
$C=10$ just before arriving at the resonance. In general a closer
spacing makes resonance capture more difficult: if the typical spacing
were $C<10$ the $\mu_\mathrm{min}$ of Equation~\ref{eqn:mumin} would
increase, making resonant capture of low-mass planets even more
unlikely. However, we see from Equation~\ref{eqn:mumin} that even
Jupiter mass planets around sun-like stars
($\mu_\mathrm{Jupiter}\simeq 9.5\times 10^{-4}$) at 10~AU lie above
$\mu_\mathrm{min}$ for the 2:1 and 3:2 resonances, and this condition
only becomes easier to satisfy closer to the star.  Thus the smaller
$C$ does not preclude resonance capture for large planets.

\section{Summary}\label{sec:summary}

Given a population of young planets embedded in a protoplanetary disk,
we derived a typical planet eccentricity as a function of planet mass
by comparing the rates of eccentricity excitation due to stirring from
nearby planets and eccentricity damping by the disk. We combined this
with disk migration rate calculations to show that in systems similar
to the early solar system, only planets $\sim$Neptune-sized or larger
are likely to both migrate slowly enough and have eccentricities small
enough for efficient capture into first-order resonance. This appears
broadly consistent with observational data on multiplanet systems,
which show that for planet pairs with a heavier exterior planet, the
majority of those in or near first-order resonance have total mass
$\gtrsim$$2M_\mathrm{Neptune}$. The agreement suggests the observed
rarity of resonant pairs around sun-like stars may arise simply
because it is too difficult for planets less massive than
$\sim$$M_\mathrm{Neptune}$ to capture into resonance in the first
place. Likewise, our results support a scenario where formation and
resonant interactions of many members of the known multiple exoplanet
systems occurred outside the ice line. We expect many pairs that fail
to capture into resonance to continue convergent migration, pass one
another, and begin divergent migration: we find direct collisions
should be rare in close approaches between planets. A significant
fraction of pairs with the lighter planet exterior could therefore
have originally been convergently migrating. For systems around M
stars, we predict that a higher fraction of planets pairs will be
resonant, as the lower mass limit for efficient resonance capture
decreases to $\sim$$M_\mathrm{Earth}$.

Ongoing and future planet searches such as HAT, K2, KELT, SPECULOOS,
and TESS will greatly increase the population of known multiplanet
systems and the range of host stellar types, providing an excellent
sample with which to confront and refine our theory. \hfill\\

\noindent {\em Acknowledgments.} We gratefully acknowledge support
from NASA grants NNX15AK23G and NNX15AM35G. We thank Re'em Sari for
pointing out Rafikov's work on planet-disk torques and gap
opening. This research has made use of the NASA Exoplanet Archive,
which is operated by the California Institute of Technology under
contract with the National Aeronautics and Space Administration under
the Exoplanet Exploration Program. The final stages of writing took
place partially at the Aspen Center for Physics, which is supported by
NSF grant PHY-1066293.

\end{document}